\documentclass[12pt]{article}%
\usepackage{amsfonts}
\usepackage{epsfig,amssymb,euscript}
\usepackage{amsmath,amscd}
\usepackage{amsmath}
\usepackage{amssymb}
\usepackage{graphicx}%
\numberwithin{equation}{section}
\newcommand{\be}{\begin{eqnarray}}
\newcommand{\ee}{\end{eqnarray}}
\newcommand{\bea}{\begin{eqnarray}}
\newcommand{\eea}{\end{eqnarray}}
\newcommand{\ba}{\begin{array}}
\newcommand{\ea}{\end{array}}

\begin{document}
\begin{titlepage}
\begin{flushright}
hep-th/yymmnnn\\
\end{flushright}
\vspace{.3cm}
\begin{center}
\renewcommand{\thefootnote}{\fnsymbol{footnote}}
{\Large{\bf Five Dimensional Non-Supersymmetric Black Holes and Strings}}
\vskip1cm
\vskip 1.3cm
J. B. Gutowski$^1$  and W. A. Sabra $^2$
\vskip 1cm
{\small{\it
$^1$DAMTP, Centre for Mathematical Sciences\\
University of Cambridge\\
Wilberforce Road, Cambridge, CB3 0WA, UK\\}}
\vskip .6cm {\small{\it
$^2$ Centre for Advanced Mathematical Sciences and Physics Department\\
American University of Beirut\\ Lebanon  \\}}
\end{center}
\bigskip
\begin{center}
{\bf Abstract}
\end{center}
We study non-supersymmetric solutions of five dimensional $N=2$ supergravity theories
coupled
to an arbitrary number of abelian vector multiplets. The solutions constructed can be
considered
as deformations of known supersymmetric black hole and string solutions.
General constraints coming from the analysis of the equations of motion are derived.
These represent explicit conditions on the charges of the black holes and strings.
The constraints are analyzed for theories where the scalar manifolds are symmetric spaces
and explicit solutions are constructed in cases where the prepotential of the theory
factorizes into a linear and a quadratic term.
\end{titlepage}

\section{Introduction}

In recent years great effort has been devoted towards the study and
classification of supersymmetric solutions of supergravity theories. In
particular, in five dimensions, it has been possible to find a complete
classification of supersymmetric solutions with various fractions of
supersymmetry \cite{susy}. The research in this domain is motivated by the
fact that supersymmetric gravitational solutions play an important role in our
understanding of the microscopic origin of entropy, stringy duality symmetries
and the conjectured AdS/CFT\ correspondence. Though much is known about the
structure of solutions preserving fractions of supersymmetry, the same can not
be said about solutions breaking all of the supersymmetries. In this paper we
are mainly interested in finding non-rotating non-supersymmetric black hole
and string solutions in five dimensional supergravity coupled to abelian
vector multiplets. Non supersymmetric solutions for these theories were first
considered in \cite{sabnonext} where an explicit solution was found for the
so-called $STU$ model with three independent electric charges. Our present
work can be considered as an elaboration and an extension of the results found
in \cite{sabnonext}. We will find non-supersymmetric electrically charged
black holes for the gauged and ungauged theories. We will also find
magnetically charged string solutions in the ungauged theories.

We organize our paper as follows. Section 2 contains a summary of the basic
notions of $N=2$, $D=5$ supergravity and very special geometry which will be
used in our subsequent analysis. In section 3 non-supersymmetric solutions are
derived for all models of gauged and ungauged $N=2$, $D=5$ supergravity
models, both the scalars and the gauge fields are expressed in terms of
harmonic functions and the equations of motion are reduced to one constraint
on these harmonic functions. Section 4 contains a similar analysis for the
non-supersymmetric black string solutions of the ungauged theories. Section 5
contains a discussion on models with symmetric scalar manifolds where the
constraints simplify and an explicit analysis is given for solutions where the
prepotential factorizes into a linear and a quadratic term. The known
solutions of the so called $STU$\ model are discussed within our analysis.
Section 6 summarizes our results.

\section{$N=2$ Supergravity theory}

Here we review some of the basics of the theories of five dimensional $N=2$
supergravity coupled to abelian vectormultiplets. Such theories were first
constructed in \cite{GST} to which the reader can be referred for detailed
discussion. The bosonic action of the theory in terms of the so-called very
special geometry was later given in \cite{wp}. A large class of the $N=2,$
$D=5$ models are obtained from the compactification of eleven dimensional
supergravity, the low energy limit of M-theory, on a Calabi-Yau threefold
\cite{cs}. The bosonic action of the $N=2$ ungauged supergravity coupled to
abelian vector multiplets can be written as
\begin{equation}
S={\frac{1}{16\pi G}}\int\left(  {}R\star1-G_{IJ}\left(  F^{I}\wedge\ast
F^{J}+dX^{I}\wedge\star dX^{J}\right)  -{\frac{1}{6}}C_{IJK}F^{I}\wedge
F^{J}\wedge A^{K}\right)  \label{act}%
\end{equation}
where $F^{I}=dA^{I}$, where $A^{I}$ are the $1$-forms representing the $n$
Abelian gauge fields. In our analysis our metric has signature $\left(
-,+,+,+,+\right)  .$ The scalars $X^{I}$ are constrained by the condition%

\begin{equation}
\mathcal{V}(X)=\frac{1}{6}\ C_{IJK}X^{I}X^{J}X^{K}=X_{I}X^{I}=1. \label{cubic}%
\end{equation}
and thus can be regarded as being functions of $n-1$ unconstrained scalars
$\phi^{i}$. The coupling $G_{IJ}$ depends on the scalars via
\begin{equation}
G_{IJ}={\frac{9}{2}}X_{I}X_{J}-{\frac{1}{2}}C_{IJK}X^{K}.
\end{equation}
Contracting $G_{IJ}$ with $X^{J}$and $\partial_{i}X^{J}$, we arrive at the
following useful equations
\begin{equation}
G_{IJ}X^{J}={\frac{3}{2}}X_{I}\,,\qquad G_{IJ}\partial_{i}X^{J}=-{\frac{3}{2}%
}\partial_{i}X_{I}\,.
\end{equation}
where $\partial_{i}=\frac{\partial}{\partial\phi^{i}}.$ The bosonic part of
the action of the corresponding $U(1)$-gauged supergravity is given by
(\ref{act}) with an additional potential term $\chi^{2}\mathcal{U},$where the
scalar potential $\mathcal{U}$ can be written as
\begin{equation}
\mathcal{U}=9V_{I}V_{J}(X^{I}X^{J}-{\frac{1}{2}}G^{IJ})
\end{equation}
where $V_{I}$ are constants. The scalar equations in the $U(1)$-gauged theory
can be written as%

\begin{align}
&  -\nabla^{\alpha}\nabla_{\alpha}X_{I}+({\frac{1}{6}}C_{MNI}-{\frac{1}{2}%
}X_{I}C_{MNJ}X^{J})\nabla_{\alpha}X^{M}\nabla^{\alpha}X^{N}%
\nonumber\label{scaleq}\\
&  -{\frac{1}{2}}\left(  X_{M}X^{P}C_{NPI}-{\frac{1}{6}}C_{MNI}-6X_{I}%
X_{M}X_{N}+{\frac{1}{6}}X_{I}C_{MNJ}X^{J}\right)  F^{M}{}_{\beta_{1}\beta_{2}%
}F^{N\beta_{1}\beta_{2}}\nonumber\\
&  -3\chi^{2}V_{M}V_{N}\left(  {\frac{1}{2}}G^{ML}G^{NP}C_{LPI}+X_{I}%
(G^{MN}-2X^{M}X^{N})\right)  =0.
\end{align}
The Einstein equations are
\begin{equation}
R_{\mu\nu}=G_{IJ}\left(  F^{I}{}_{\mu\lambda}F^{J}{}_{\nu}{}^{\lambda}%
+\nabla_{\mu}X^{I}\nabla_{\nu}X^{J}-\frac{1}{6}g_{\mu\nu}F^{I}{}_{\rho\sigma
}F^{J\rho\sigma}\right)  -\frac{2}{3}\chi^{2}g_{\mu\nu}\mathcal{U} \label{eem}%
\end{equation}
and the Maxwell gauge equations are given by
\begin{equation}
d\left(  G_{IJ}\star F^{J}\right)  =-{\frac{1}{4}}C_{IJK}F^{J}\wedge F^{K}.
\label{geom}%
\end{equation}

\section{Non-Supersymmetric Black Holes}

In this section we construct a class of non-supersymmetric black holes in both
the gauged and the ungauged supergravity theories, we consider the gauged
theory first. We take the following ansatz for the metric:
\begin{equation}
ds^{2}=-e^{-4A}fdt^{2}+e^{2A}\left(  \frac{dr^{2}}{f}+r^{2}d\Omega_{3,k}%
^{2}\right)  \
\end{equation}
where $A=A(r)$, $f=f(r)$ and
\begin{equation}
d\Omega_{3,k}^{2}=%
\begin{cases}
d\xi^{2}+\sin^{2}\xi(d\theta^{2}+\sin^{2}\theta d\phi^{2})\qquad & k=1\\
d\xi^{2}+\xi^{2}(d\theta^{2}+\sin^{2}\theta d\phi^{2})\qquad & k=0\\
d\xi^{2}+\sinh^{2}\xi(d\theta^{2}+\sin^{2}\theta d\phi^{2})\qquad & k=-1
\end{cases}
\end{equation}
corresponds to the metric on $\mathbf{S}^{3}$, $\mathbf{R}^{3}$ or
$\mathbf{H}^{3}$ according as $k=1$, $k=0$ or $k=-1$.

The non-vanishing Ricci tensor components are given by%

\begin{align}
R_{tt}  &  =-\frac{f}{2r}e^{-6A}\left(  4A^{\prime\prime}rf+4f^{\prime
}A^{\prime}r-rf^{\prime\prime}+12A^{\prime}f-3f^{\prime}\right)  ,\nonumber\\
R_{rr}  &  =-\left(  A^{\prime\prime}+6A^{\prime2}+3\frac{A^{\prime}}%
{r}\right)  -\frac{1}{2rf}\left(  -4f^{\prime}A^{\prime}r+rf^{\prime\prime
}+3f^{\prime}\right)  ,\nonumber\\
R_{\xi\xi}  &  =-f\left(  3A^{\prime}r+A^{\prime\prime}r^{2}+2\right)
-f^{\prime}\left(  A^{\prime}r^{2}+r\right)  +2k
\end{align}
and
\begin{equation}
R_{\theta\theta}=%
\begin{cases}
\sin^{2}\xi\ R_{\xi\xi}\qquad & k=1\\
\xi^{2}\ R_{\xi\xi}\qquad & k=0\\
\sinh^{2}\xi\ R_{\xi\xi}\qquad & k=-1
\end{cases}
\end{equation}%
\begin{equation}
R_{\phi\phi}=%
\begin{cases}
\sin^{2}\xi\sin^{2}\theta\ R_{\xi\xi}\qquad & k=1\\
\xi^{2}\sin^{2}\theta\ R_{\xi\xi}\qquad & k=0\\
\sinh^{2}\xi\sin^{2}\theta\ R_{\xi\xi}\qquad & k=-1
\end{cases}
\end{equation}
The prime denotes differentiation with respect to the radial coordinate $r.$

To proceed, we assume that the only non-vanishing component of the gauge field
strengths is $F^{I}{}_{tr}=F^{I}{}_{tr}(r)$, and that the scalars $X^{I}$
depend only on $r$. We take the following as an ansatz for the gauge fields
{\footnote{Note that in the case $k=-1$, we have complexified the gauge field
strengths. Such solutions, strictly speaking, are not non-extremal solutions
of the standard $N=2$ supergravity. Rather, they are solutions of a modified
theory, with a sign change in the Maxwell term in the action.}}%

\begin{align}
F_{rt}^{I}  &  =\frac{1}{2}e^{-4A}G^{IJ}\partial_{r}\tilde{H}_{J},\text{
\ \ for \ \ }k=0,\text{\ }1,\\
F_{rt}^{I}  &  =\frac{i}{2}e^{-4A}G^{IJ}\partial_{r}\tilde{H}_{J},\text{
\ \ for \ \ }k=-\text{\ }1,
\end{align}
where $\tilde{H}_{I}$ constitute a set of harmonic functions $\tilde{H}%
_{I}=\tilde{h}_{I}+\frac{\tilde{q}_{I}}{r^{2}}.$ Both the gauge field
equations and the Bianchi identities hold without further constraint.

Next consider the Einstein equations; these are equivalent to
\begin{equation}
f^{\prime\prime}+{\frac{7}{r}}f^{\prime}+{\frac{8}{r^{2}}}f-{\frac{8k}{r^{2}}%
}=-36\chi^{2}e^{2A}V_{I}V_{J}\left(  {\frac{1}{2}}G^{IJ}-X^{I}X^{J}\right)
\label{ein1}%
\end{equation}
and
\begin{equation}
G_{IJ}F_{rt}^{I}F_{rt}^{J}=e^{-4U}\left(  -3f^{\prime}A^{\prime}%
-3fA^{\prime\prime}+\frac{f^{\prime\prime}}{2}-{\frac{9fA^{\prime}}{r}}%
+{\frac{f^{\prime}}{2r}}+{\frac{2(k-f)}{r^{2}}}\right)  \label{ein2}%
\end{equation}
and
\begin{equation}
G_{IJ}\partial_{r}X^{I}\partial_{r}X^{J}=-\left(  3A^{\prime\prime}+{\frac
{9}{r}}A^{\prime}+6A^{\prime}{}^{2}\right)  . \label{ein3}%
\end{equation}

To satisfy these constraints, we adopt the same ansatz for the scalars as for
the ungauged supersymmetric black hole solutions \cite{waf1}:%

\begin{equation}
X_{I}=\frac{1}{3}e^{-2A}H_{I}(r).
\end{equation}
where $H_{I}(r)$ are harmonic functions. This constraint is sufficient to
ensure that ({\ref{ein3}}) is satisfied.

We set
\begin{equation}
H_{I}=\delta V_{I}+{\frac{q_{I}}{r^{2}}} \label{harmfun1}%
\end{equation}
where $\delta$ is a non-zero constant. Then ({\ref{ein1}}) can be rewritten
as
\begin{equation}
f^{\prime\prime}+{\frac{7}{r}}f^{\prime}+{\frac{8}{r^{2}}}f-{\frac{8k}{r^{2}}%
}={\frac{9\chi^{2}}{\delta^{2}}}\big((r^{2}e^{6A})^{\prime\prime}+{\frac{7}%
{r}}(r^{2}e^{6A})^{\prime}+8e^{6A}\big)
\end{equation}

This equation has
\begin{equation}
f=k-{\frac{\mu}{r^{2}}}+{\frac{9\chi^{2}}{\delta^{2}}}r^{2}e^{6A} \label{fsol}%
\end{equation}
as a solution. The remaining condition ({\ref{ein2}}) from the Einstein
equations is then equivalent to%

\begin{equation}
G^{IJ}S_{IJ}=0 \label{contract1}%
\end{equation}
where
\begin{align}
S_{IJ}  &  =\left(  {\tilde{q}}_{I}{\tilde{q}}_{J}-kq_{I}q_{J}\right)
-{\frac{1}{2}}\mu\delta(q_{I}V_{J}+q_{J}V_{I}),\text{ \ for }k=0,1,\nonumber\\
S_{IJ}  &  =-\left(  {\tilde{q}}_{I}{\tilde{q}}_{J}-q_{I}q_{J}\right)
-{\frac{1}{2}}\mu\delta(q_{I}V_{J}+q_{J}V_{I}),\text{ \ for }k=-1,
\end{align}
Lastly, we consider the scalar equation ({\ref{scaleq}}). It is
straightforward but tedious to show that constraint ({\ref{contract1}}) and
the scalar equations are equivalent to the constraint%

\begin{equation}
C_{MNI}G^{ML}G^{NT}S_{LT}+8X^{M}S_{MI}-12X_{I}X^{M}X^{N}S_{MN}=0.
\label{harmcon1}%
\end{equation}
We note that if we contract (\ref{harmcon1}) with $X^{I},$ then equation
(\ref{contract1}) is obtained

For supersymmetric black holes with event horizon topology $S^{3}$, we take
$\mu=0$, $k=1$ and $H_{I}=\tilde{H}_{I}$, and ({\ref{harmcon1}}) is satisfied
with $S_{IJ}=0$. However, for the deformed solutions with $\mu\neq0$, if
$S_{IJ}=0$ for all $I,J$ then it is straightforward to show that there must
exist constants $\alpha$, $\beta$ such that
\begin{equation}
{\tilde{q}}_{I}=\alpha q_{I}+\beta V_{I}%
\end{equation}
and furthermore $q_{I}$ and $V_{I}$ must be linearly dependent. In order to
find solutions for which the charges are not so strongly constrained, instead
of solving $S_{IJ}=0$ for all $I,J$, one must solve the weaker condition given
by ({\ref{harmcon1}}).

Finally, the non-supersymmetric black hole solutions of the ungauged theory
are obtained by setting $\chi=0$ and $k=1$ throughout the gauged solution. One
minor subtlety is that for the gauged solutions, the asymptotic values of the
harmonic functions $H_{I}$ given in ({\ref{harmfun1}}) are fixed (up to an
overall scale) in terms of the constants $V_{I}$ which appear in the
construction of the theory. However, for the ungauged solutions, the constants
$V_{I}$ appearing in ({\ref{harmfun1}}) are arbitrary. Setting $\mu=0,$ one
recovers the supersymmetric black hole solutions presented in \cite{waf1, bcs}.

\section{Non-Supersymmetric Magnetic Strings}

In this section we construct non-supersymmetric black string solutions of the
ungauged theory. The metric is given by%

\begin{equation}
ds^{2}=e^{-2B}\left(  -fdt^{2}+dz^{2}\right)  +e^{4B}\left(  \frac{1}{f}%
dr^{2}+r^{2}d\theta^{2}+r^{2}\sin^{2}\theta d\phi^{2}\right)
\end{equation}
where $B=B(r)$, $f=f(r)$. The non-vanishing components of the Ricci tensor are
given by%

\begin{align}
R_{tt}  &  =-e^{-6B}f\left(  f^{\prime}B^{\prime}-\frac{f^{\prime}}{r}%
-\frac{f^{\prime\prime}}{2}\right)  -e^{-6B}f^{2}\left(  B^{\prime\prime
}+\frac{2B^{\prime}}{r}\right)  ,\nonumber\\
R_{zz}  &  =e^{-6B}f^{\prime}B^{\prime}+fe^{-6B}\left(  B^{\prime\prime}%
+\frac{2B^{\prime}}{r}\right)  ,\nonumber\\
R_{rr}  &  =-\left(  2B^{\prime\prime}+6B^{\prime2}-\frac{f^{\prime}B^{\prime
}}{f}+\frac{f^{\prime\prime}}{2f}+\frac{4B^{\prime}}{r}+\frac{f^{\prime}}%
{rf}\right)  ,\nonumber\\
R_{\theta\theta}  &  =-f\left(  4B^{\prime}r+2B^{\prime\prime}r^{2}+1\right)
-rf^{\prime}\left(  2B^{\prime}r+1\right)  +1,\nonumber\\
R_{\phi\phi}  &  =R_{\theta\theta}\sin^{2}\theta. \label{meq}%
\end{align}
We also assume that the scalars $X^{I}$ depend only on $r$, and that the only
non-zero components of the gauge field strengths are given by
\begin{equation}
F_{\theta\phi}^{I}=\alpha^{I}\sin\theta
\end{equation}
for constant $\alpha^{I}$. With these choices, the gauge field equations and
Bianchi identities hold without further constraint. The Einstein equations
(taking $\chi=0$) then fix
\begin{equation}
f=1-\frac{\mu}{r}%
\end{equation}
together with the constraints
\begin{equation}
\frac{e^{-4B}}{r^{4}}G_{IJ}\alpha^{I}\alpha^{J}=-3\left(  f^{\prime}B^{\prime
}+fB^{\prime\prime}+\frac{2B^{\prime}f}{r}\right)  , \label{einb1}%
\end{equation}%
\begin{equation}
G_{IJ}\partial_{r}X^{J}\partial_{r}X^{I}=-3\left(  B^{\prime\prime}%
+2B^{\prime2}+\frac{2B^{\prime}}{r}\right)  . \label{einb2}%
\end{equation}
To satisfy the constraint ({\ref{einb2}}) we set%

\begin{equation}
X^{I}=e^{-2B}H^{I}%
\end{equation}
where
\begin{equation}
H^{I}=h^{I}+{\frac{q^{I}}{r}}%
\end{equation}
are harmonic functions. Then ({\ref{einb1}}) can be rewritten as
\begin{equation}
G_{IJ}U^{IJ}=0 \label{magharm1}%
\end{equation}
where
\begin{equation}
U^{IJ}=\alpha^{I}\alpha^{J}-q^{I}q^{J}-{\frac{1}{2}}\mu(h^{I}q^{J}+h^{J}%
q^{I}).
\end{equation}
Finally, consider the scalar equations ({\ref{scaleq}}) (with $\chi=0$). It is
straightforward to show that the scalar equations, together with
({\ref{magharm1}}) are equivalent to
\begin{equation}
\left(  X_{M}X^{P}C_{NPI}-{\frac{1}{6}}C_{MNI}-{\frac{9}{2}}X_{I}X_{M}%
X_{N}\right)  U^{MN}=0. \label{stringcon}%
\end{equation}

Again if we contract (\ref{stringcon}) with $X^{I}$ the condition
(\ref{magharm1}) is obtained. Note that (\ref{stringcon}) can be rewritten
entirely in terms of the harmonic functions $H^{I}$ as
\begin{align}
U^{MN} \bigg( C_{MM_{1}M_{2}}C_{INM_{3}}C_{M_{4}M_{5}M_{6}}-{\frac{1}{6}
}C_{MNI}C_{M_{1}M_{2}M_{3}}C_{M_{4}M_{5}M_{6}}\nonumber\\
-{\frac{3}{4}}C_{IM_{1}M_{2} }C_{MM_{3}M_{4}}C_{NM_{5}M_{6}} \bigg) H^{M_{1}%
}H^{M_{2}}H^{M_{3}}H^{M_{4}}H^{M_{5}}H^{M_{6}} =0. \label{harms}%
\end{align}

Clearly, one way to satisfy ({\ref{harms}}) is to set $U^{MN}=0$ for all
$M,N$; however just as in the case of the black holes, this constraint is too
restrictive on the charges. Finally, for $\mu=0,$ one obtains the
supersymmetric magnetic strings constructed in \cite{waf2}

\section{Explicit Solutions}

In this section, we shall construct solutions for the models related to Jordan
algebras, i.e., models where the scalar manifold is a symmetric space. These
theories were first constructed by Gunaydin, Sierra and Townsend \cite{GST}
where it was shown that $\mathcal{V}$ \ are in one-to-one correspondence with
the norm forms of Euclidean (formally real) Jordan algebras $J$ of degree 3.
The target spaces take the form
\begin{equation}
\mathcal{M}=\frac{\mathrm{Str}_{0}\left(  J\right)  }{\mathrm{Aut}\left(
J\right)  }.
\end{equation}
Here $\mathrm{Str}_{0}\left(  J\right)  $ denotes the invariance group of the
norm (reduced structure group) of the Jordan algebra $J$ and $\mathrm{Aut}%
\left(  J\right)  $ is its automorphism group. Non-simple Jordan algebras of
degree three are of the form $\mathbb{R\oplus}\Gamma_{n}$ , where $\Gamma_{n}$
is the Jordan algebra associated with a quadratic form. The corresponding
symmetric scalar manifolds are%

\begin{equation}
\mathcal{M}=SO(1,1)\times\frac{SO(n-1,1)}{SO(n-1)}. \label{scalman}%
\end{equation}

In this case, $\mathcal{V}(X)$ is factorizable into a linear times a quadratic
form in $(n-1)$ scalars, which for the positivity of the kinetic terms in the
Lagrangian, must have a Minkowski metric. For Simple Euclidean Jordan algebras
$\mathfrak{h}_{3}(\mathbb{A)}$ generated by $3\times3$ Hermitian matrices over
the four division algebras $\mathbb{A}=\mathbb{R}$, $\mathbb{C}$, $\mathbb{H}%
$, $\mathbb{O}$, the corresponding spaces $\mathcal{M}$ are, respectively:
\[
\mathcal{M=}\dfrac{\mathrm{SL}\left(  3,\mathbb{R}\right)  }{\mathrm{SO}%
\left(  3\right)  },\text{ \ \ }\dfrac{\mathrm{SL}\left(  3,\mathbb{C}\right)
}{\mathrm{SU}\left(  3\right)  },\text{ \ }\dfrac{\mathrm{SU}^{\ast}\left(
6\right)  }{\mathrm{USp}\left(  6\right)  },\text{ \ }\dfrac{\mathrm{E}%
_{6(-26)}}{\mathrm{F}_{4}}.\text{ }%
\]
For the simple Jordan algebras \cite{jordan}, an element for the four families
$\mathfrak{h}_{3}^{\mathbb{A}}$ can be written in the form%

\begin{equation}
L=\left(
\begin{array}
[c]{ccc}%
\alpha & z^{\ast} & y^{\ast}\\
z & \beta & x\\
y & x^{\ast} & \gamma
\end{array}
\right)
\end{equation}
where ($\alpha,\beta,\gamma)\in\mathbb{R}$ and $(x,y,z)\in\mathbb{A}$. The
cubic norm $\mathcal{V}$ is given by
\begin{equation}
\mathcal{V}=\det L=\alpha\beta\gamma-\left(  \alpha\left\vert {x}\right\vert
^{2}+\beta\left\vert {y}\right\vert ^{2}+\gamma\left\vert {z}\right\vert
^{2}\right)  +2\operatorname{Re}\left(  xyz\right)
\end{equation}
In all of these cases, the following constraints hold:
\begin{align}
C^{IJK}  &  =\delta^{II^{\prime}}\delta^{JJ^{\prime}}\delta^{KK^{\prime}%
}C_{I^{\prime}J^{\prime}K^{\prime}}\,,\\
C_{IJK}C_{J^{\prime}\left(  LM\right.  }C_{\left.  PQ\right)  K^{\prime}%
}\delta^{JJ^{\prime}}\delta^{KK^{\prime}}  &  =\frac{4}{3}\delta_{I\left(
L\right.  }C_{\left.  MPQ\right)  },\\
X^{I}  &  ={\frac{9}{2}}C^{IJK}X_{J}X_{K}\\
G^{IJ}  &  =2X^{I}X^{J}-6C^{IJK}X_{K}%
\end{align}

\bigskip In the case of the black hole solutions, the constraint
({\ref{harmcon1}}) can then be rewritten as
\begin{equation}
\left(  C^{IMN}-6X^{M}C^{IJN}X_{J}+X^{I}X^{M}X^{N}\right)  S_{MN}=0
\label{symcon1}%
\end{equation}

or equivalently
\begin{align}
S_{MN}\bigg( {\frac{1}{36}}C^{M_{1}M_{2}M_{3}}C^{M_{4}M_{5}M_{6}}
C^{LMN}-{\frac{1}{6}}C^{M_{1}M_{2}M_{3}}C^{NLM_{4}}C^{MM_{5}M_{6}}\nonumber\\
+{\frac{1}{8}}C^{LM_{1}M_{2}}C^{MM_{3}M_{4}}C^{NM_{5}M_{6}}\bigg) H_{M_{1}%
}H_{M_{2}}H_{M_{3}}H_{M_{4}}H_{M_{5}}H_{M_{6}} =0. \label{harmbh}%
\end{align}
Observe that this equation is identical (up to a trivial raising and lowering
of indices) to that found for the black string solutions in ({\ref{harms}}).
This is also expected because of the duality symmetry discussed in
\cite{wafkz}. Hence, it suffices to solve the equation ({\ref{harmbh}}), (or
equivalently ({\ref{symcon1}})). These equations can be simplified slightly to
give
\begin{equation}
\left(  {\frac{1}{2}}C_{IM_{1}M_{2}}C^{M_{1}MN}C^{M_{2}N_{1}N_{2}}-\delta
_{I}^{M}C^{NN_{1}N_{2}}\right)  S_{MN}H_{N_{1}}H_{N_{2}}=0.
\end{equation}
In principle, the constraints on the charges can be obtained by expanding this
equation in powers of $r$; however these constraints are highly non-linear and
in general they do not appear tractable.

To proceed, we consider the case when the pre-potential ${\mathcal{V}}$
factorizes into a linear times a quadratic form as
\begin{equation}
{\mathcal{V}}={\frac{1}{2}}X^{1}\left(  \eta_{ab}X^{a}X^{b}\right)  ,\quad
a,b=2,\dots,n
\end{equation}
and $\eta_{ab}$ is a Minkowski metric on $\mathbb{R}^{1,n-2}$..

Then we note the useful identities
\begin{align}
\eta_{ab}X^{b}  &  =9X_{a}X_{1},\quad X^{1}={\frac{9}{2}}\eta^{ab}X_{a}%
X_{b},\quad X^{a}=9X_{1}\eta^{ab}X_{b}\nonumber\\
X^{1}X_{1}  &  ={\frac{1}{3}},\quad X^{a}X_{a}={\frac{2}{3}},\quad\eta
^{ab}X_{b}={\frac{1}{3}}X^{1}X^{a}.
\end{align}
It is then straightforward to show that the constraints ({\ref{symcon1}}) give
the two conditions
\begin{equation}
S_{11}=0, \label{ssone}%
\end{equation}
and
\begin{equation}
2X^{c}\eta^{ab}S_{bc}-X^{a}\eta^{bc}S_{bc}=0. \label{secondcon}%
\end{equation}
Note that the components $S_{1a}$ are not constrained by ({\ref{symcon1}}).
The constraint $S_{11}=0$ is equivalent to%
\begin{align}
\tilde{q}_{1}  &  =\sqrt{\mu\delta q_{1}V_{1}},\text{ \ for }k=0,\nonumber\\
\tilde{q}_{1}  &  =\sqrt{q_{1}^{2}+\mu\delta q_{1}V_{1}},\text{ \ for
}k=1,\nonumber\\
\tilde{q}_{1}  &  =\sqrt{q_{1}^{2}-\mu\delta q_{1}V_{1}},\text{ \ for
}k=-1,\text{\ }%
\end{align}
and ({\ref{secondcon}}) is equivalent to%

\begin{equation}
\left(  \delta V_{d}+{\frac{q_{d}}{r^{2}}}\right)  \left(  2\eta^{cd}\eta
^{ab}-\eta^{ad}\eta^{bc}\right)  S_{bc}=0. \label{base}%
\end{equation}
This gives two equations (for $k=0,1)$%

\begin{align}
2V^{b}{\tilde{q}}_{b}{\tilde{q}}_{a}-(2kV^{b}q_{b}+\mu\delta V^{b}V_{b}%
)q_{a}-({\tilde{q}}^{b}{\tilde{q}}_{b}-kq^{b}q_{b})V_{a}  &  =0\nonumber\\
2q^{b}{\tilde{q}}_{b}{\tilde{q}}_{a}-({\tilde{q}}^{b}{\tilde{q}}_{b}%
+kq^{b}q_{b})q_{a}-\mu\delta q^{b}q_{b}V_{a}  &  =0 \label{nonlin}%
\end{align}
where $q^{a}=\eta^{ab}q_{b}$, $V^{a}=\eta^{ab}V_{b}$, $\tilde{q}^{a}=\eta
^{ab}\tilde{q}_{b}$.

There are a number of cases to consider. In the first case, there exist
$\lambda,\sigma$ such that
\begin{equation}
\tilde{q}_{a}=\lambda q_{a}+\sigma V_{a},
\end{equation}
then ({\ref{nonlin}}) can be rewritten as%

\begin{align}
\bigg( 2\left(  \lambda^{2}-k \right)  V^{b}q_{b}+\left(  2\lambda\sigma
-\mu\delta\right)  V^{b} V_{b}\bigg) q_{a}+\bigg( \sigma^{2}V^{b}V_{b}-\left(
\lambda^{2}-k \right)  q^{b} q_{b}\bigg) V_{a}  &  =0\nonumber\\
\bigg( \left(  \lambda^{2}-k \right)  q^{b}q_{b}-\sigma^{2}V^{b}%
V_{b}\bigg) q_{a}+\bigg( \left(  2\lambda\sigma-\mu\delta\right)  q^{b}%
q_{b}+2\sigma^{2}V^{b}q_{b}\bigg) V_{a}  &  =0
\end{align}
There are then two sub-cases.

{(i)} $\sigma^{2}V_{b}V^{b}-(\lambda^{2}-k)q^{b}q_{b}\neq0$. Then there exists
$\theta$ such that $q_{a}=\theta V_{a}$ for all $a$, where%

\begin{align}
\theta\neq0, \quad V^{a}V_{a} \neq0, \quad\theta^{2} (\lambda^{2}
-k)-\sigma^{2}  &  \neq0\nonumber\\
\theta^{2}(\lambda^{2}-k)+\theta(2\lambda\sigma-\mu\delta)+\sigma^{2}  &  = 0
\end{align}

{(ii)} $\sigma^{2}V_{b}V^{b}-(\lambda^{2}-k)q^{b}q_{b}=0$. There are then four possibilities:

\begin{enumerate}
\item $q_{a}=0$ for all $a$ with $\sigma=0$ and $V^{b}V_{b}\neq0$.

\item $V_{a}=0$ for all $a$ with $\lambda^{2}=k$, $2\lambda\sigma-\mu
\delta\neq0$, $q^{b}q_{b}\neq0$.

\item $\sigma\neq0$ with
\begin{equation}
q^{b}V_{b}=-{\frac{(2\lambda\sigma-\mu\delta)}{2\sigma^{2}}}q^{b}q_{b},\qquad
V^{b}V_{b}={\frac{(\lambda^{2}-k)}{\sigma^{2}}}q^{b}q_{b}. \label{kpluszero}%
\end{equation}

\item $\sigma=0$ with $q^{b}q_{b}=0$ and $V^{b}V_{b}={\frac{2(\lambda^{2}%
-k)}{\mu\delta}}V^{b}q_{b}$.
\end{enumerate}

One can also consider the case where $V^{b}{\tilde{q}}_{b}=q^{b}{\tilde{q}%
}_{b}=0$. There are then two sub-cases:

\begin{enumerate}
\item If $q^{b}q_{b}\neq0$ then there exists $\lambda$ such that
$V_{a}=\lambda q_{a}$ for all $a$. $\lambda$ is then fixed by $(k+\lambda
\mu\delta)q^{b}q_{b}+\tilde{q}^{b}\tilde{q}_{b}=0$,

\item $q_{a}=0$ for all $a$, and $\tilde{q}^{a}\tilde{q}_{a}=0$. (Note that we
cannot have both $q_{a}=0$ and $V_{a}=0$ for all $a$, as this would imply
$X_{a}=0$ for all $a$, in contradiction to the constraint $X^{a}X_{a}%
={\frac{2}{3}}$).
\end{enumerate}

Similarly for the case of $k=-1,$ then equation (\ref{secondcon}) gives two equations%

\begin{align}
2V^{b}{\tilde{q}}_{b}{\tilde{q}}_{a}+(-2V^{b}q_{b}+\mu\delta V^{b}V_{b}%
)q_{a}-(\tilde{q}^{b}\tilde{q}_{b}-q^{b}q_{b})V_{a}  &  =0\nonumber\\
2q^{b}{\tilde{q}}_{b}{\tilde{q}}_{a}+(-\tilde{q}^{b}\tilde{q}_{b}-q^{b}%
q_{b})q_{a}+\mu\delta q^{b}q_{b}V_{a}  &  =0
\end{align}
and similarly there are a number of cases to consider. Again one can consider
the case when
\begin{equation}
\tilde{q}_{a}=\lambda q_{a}+\sigma V_{a},
\end{equation}

This gives
\begin{align}
\bigg( 2\left(  \lambda^{2}-1\right)  V^{b}q_{b}+\left(  \mu\delta
+2\sigma\lambda\right)  V^{b}V_{b} \bigg)q_{a}-\bigg(\left(  \lambda
^{2}-1\right)  q^{b}q_{b}-\sigma^{2}V^{b}V_{b}\bigg)V_{a}  &  =0\nonumber\\
\bigg(\left(  \lambda^{2}-1\right)  q^{b}q_{b}-\sigma^{2}V^{b}V_{b}%
\bigg)q_{a}+\bigg( 2\sigma^{2}q^{b}V_{b}+\left(  2\lambda\sigma+\mu
\delta\right)  q^{b}q_{b}\bigg) V_{a}  &  =0
\end{align}

and as for the $k=0,1$, we consider two sub-cases.

{(i)} $\left(  \lambda^{2}-1\right)  q^{b}q_{b}-\sigma^{2}V^{b}V_{b}\neq0$.
Then there exists $\theta$ such that $q_{a}=\theta V_{a}$ for all $a$, where
\begin{align}
\theta\neq0, \quad V^{a}V_{a} \neq0, \quad\theta^{2}(\lambda^{2}-1)-\sigma^{2}
\neq0\nonumber\\
\theta^{2}(\lambda^{2}-1)+\theta(2\lambda\sigma+\mu\delta)+\sigma^{2}
=0\nonumber
\end{align}

{(ii)} $\left(  \lambda^{2}-1\right)  q^{b}q_{b}-\sigma^{2}V^{b}V_{b}=0$. Then
we have four possibilities:

\begin{enumerate}
\item $q_{a}=0$ for all $a$ with $\sigma=0$ and $V^{b}V_{b}\neq0$.

\item $V_{a}=0$ for all $a$ with $\lambda^{2}=1$, $2\lambda\sigma+\mu
\delta\neq0$, $q^{b}q_{b}\neq0$.

\item $\sigma\neq0$ with
\begin{equation}
q^{b}V_{b}=-{\frac{(2\lambda\sigma+\mu\delta)}{2\sigma^{2}}}q^{b}q_{b},\qquad
V^{b}V_{b}={\frac{(\lambda^{2}-1)}{\sigma^{2}}}q^{b}q_{b}. \label{kminus}%
\end{equation}

\item $\sigma=0$ with $q^{b}q_{b}=0$ and $V^{b}V_{b}=-{\frac{2(\lambda^{2}%
-1)}{\mu\delta}}V^{b}q_{b}$.
\end{enumerate}

Also we consider the case when $V^{b}\tilde{q}_{b}=q^{b}\tilde{q}_{b}=0$. Then
there are two sub-cases:

\begin{enumerate}
\item If $q^{b}q_{b}\neq0$ then there exists $\lambda$ such that
$V_{a}=\lambda q_{a}$ for all $a$. $\lambda$ is then fixed by $\tilde{q}%
^{b}\tilde{q}_{b}+\left(  1-\lambda\mu\delta\right)  q^{b}q_{b}=0$,

\item $q_{a}=0$ for all $a$, and $\tilde{q}^{a}\tilde{q}_{a}=0$.
\end{enumerate}

\bigskip

It is instructive to see where the non-extremal $STU$ black hole solutions of
\cite{sabnonext} fit into this scheme. For the $STU$ model, we take $X^{1}=S$,
$X^{2}=T$, $X^{3}=U$ with
\begin{equation}
\eta_{ab}=\left(
\begin{array}
[c]{cc}%
0 & 1\\
1 & 0
\end{array}
\right)  .
\end{equation}
The black hole solutions with spherical horizons correspond to setting
$\chi=g$, $k=1$, $\delta=3$ with $V_{1}=V_{2}=V_{3}={\frac{1}{3}}$ and
\begin{equation}
{\tilde{q}}_{a}=\mu\sinh\beta_{a}\cosh\beta_{a},\qquad q_{a}=\mu\sinh^{2}%
\beta_{a},\quad a=2,3
\end{equation}
This solution corresponds to the case (\ref{kpluszero}) with where $\sigma
\neq0$ and $k=1$. Consider the case for which $\beta_{2}\neq\beta_{3}$, one
finds that
\[
{\tilde{q}}_{a}=\lambda q_{a}+\sigma V_{a}%
\]
for $a=2,3$, with
\begin{align}
\lambda &  ={\frac{\sinh\beta_{2}\cosh\beta_{2}-\sinh\beta_{3}\cosh\beta_{3}%
}{\sinh^{2}\beta_{2}-\sinh^{2}\beta_{3}},}\nonumber\\
\sigma &  =3\mu\left(  {\frac{\sinh\beta_{2}\sinh\beta_{3}}{\sinh^{2}\beta
_{2}-\sinh^{2}\beta_{3}}}\right)  \left(  \sinh\beta_{2}\cosh\beta_{3}%
-\sinh\beta_{3}\cosh\beta_{2}\right)  .
\end{align}
For $k=0$, we have

\begin{equation}
\tilde{q}_{a}={\mu\sinh\beta_{a},}\text{ \ \ \ \ \ }q_{a}={\mu\sinh}^{2}%
{\beta_{a},}%
\end{equation}
In this case it is easy to verify that
\[
\tilde{q}_{a}={\lambda}q_{a}+\sigma V_{a}%
\]
for  $a=2,3$, with%

\begin{equation}
{\lambda}=\frac{1}{\left(  {\sinh\beta_{2}+\sinh\beta}_{3}\right)  }%
,\qquad\sigma=3{\mu}\frac{{\sinh\beta_{3}\sinh\beta_{2}}}{{\sinh\beta
_{2}+\sinh\beta}_{3}}.
\end{equation}
This belongs to the class of solutions satisfying (\ref{kpluszero}) with
$k=0.$

For $k=-1,$ we have%
\begin{equation}
\tilde{q}_{I}=-{\mu\sinh\beta_{I}\cosh\beta_{I},}\text{ \ \ \ \ \ }q_{I}%
=-{\mu\sinh}^{2}{\beta_{I}.}%
\end{equation}
Then
\[
\tilde{q}_{a}=\lambda q_{a}+\sigma V_{a}%
\]
for $a=2,3$, with
\begin{align}
\lambda &  ={\frac{\sinh\beta_{2}\cosh\beta_{2}-\sinh\beta_{3}\cosh\beta_{3}%
}{\sinh^{2}\beta_{2}-\sinh^{2}\beta_{3}}}\nonumber\\
\sigma &  =-\left(  {\frac{3\mu\sinh\beta_{2}\sinh\beta_{3}}{\sinh^{2}%
\beta_{2}-\sinh^{2}\beta_{3}}}\right)  \left(  \sinh\beta_{2}\cosh\beta
_{3}-\sinh\beta_{3}\cosh\beta_{2}\right)
\end{align}
This belongs to the class of solutions satisfying (\ref{kminus}).

\section{Discussion}

We have constructed non-supersymmetric solutions of five dimensional $N=2$
supergravity theories coupled to an arbitrary number of abelian vector
multiplets. The solutions constructed are deformations of known supersymmetric
black hole and string solutions. The scalar fields have the same solution as
in the supersymmetric cases. However, one has to solve extra conditions
involving the various charges and the parameter $\mu.$ These conditions are
given for the black holes and black strings, respectively by (\ref{harmcon1})
and (\ref{stringcon}). However, for supergravity models with scalars living on
symmetric spaces the condition (\ref{harmcon1}) take a much simpler form given
in (\ref{symcon1}) which can also be obtained from (\ref{stringcon}) using the
duality transformation discussed in \cite{wafkz}. We have studied the
condition (\ref{symcon1}) for models where the prepotential factorizes into a
linear and a quadratic form and derived various conditions for the existence
of explicit solutions. It is of interest to find more general
non-supersymmetric solutions as deformations of known general supersymmetric
ones. Our results can be generalized to other supergravity models and in
particular to those in four dimensions. We hope to report on this in a future publication.\ 

\bigskip

\textbf{Acknowledgement:}

This work of W. S is supported in part by the National Science Foundation
under grant number PHY-0703017.

\end{document}